# Grain boundaries control lithiation of solid solution substrates in lithium metal batteries

Leonardo Shoji Aota[a], Chanwon Jung[a,b], Siyuan Zhang[a], Ömer K. Büyükuslu[c], Poonam Yadav[a], Mahander Pratap Singh[a], Xinren Chen[a], Eric Woods[a], Christina Scheu[a], Se-Ho Kim[a,d], Dierk Raabe[a], Baptiste Gault[a,e]

[a] Max-Planck-Institute for Sustainable Materials, 40237, Düsseldorf, Germany

[b] Pukyong National University, Busan, 48513, Republic of Korea

[c] GTT-Technologies, 52134, Herzogenrath, Germany

[d] Department of Materials Science and Engineering, Korea University, Seoul 02841, Republic of Korea

[e] Department of Materials, Imperial College London, London, SW7 2AZ, UK

## Abstract

The development of sustainable transportation and communication systems requires an increase in both energy density and capacity retention of Li-batteries. Using substrates forming a solid solution with body centered cubic Li enhances the cycle stability of anode-less batteries. However, it remains unclear how the substrate microstructure affects the lithiation behavior. Here, we deploy a correlative, near-atomic scale probing approach through combined ion- and electron-microscopy to examine the distribution of Li in Li-Ag diffusion couples as model system. We reveal that Li regions with over 93.8% at.% nucleate within Ag at random high angle grain boundaries, whereas grain interiors are not lithiated. We evidence the role of kinetics and mechanical constraint from the microstructure over equilibrium thermodynamics in dictating the lithiation process. The findings suggest that grain size and grain boundary character are critical to enhance the electrochemical performance of interlayers/electrodes, particularly for improving lithiation kinetics and hence reducing dendrite formation.

**Introduction**

With the increasing demand for energy storage from renewable sources[1], batteries have become vital to a transition towards a more sustainable society. Among many applications, lithium-ion batteries stand out due to their high energy density, especially suitable for electric vehicles and electronic devices[2,3]. The increasing demand for Li-ion batteries drives the improvement of their energy density, power and capacity retention while employing cheaper and more sustainable materials[1,2].

Currently, graphite anodes are dominating the commercial market. Although graphite offers stable cyclability, it exhibits a limited gravimetric capacity of only 372 mAh/g[4]. Additionally, graphite shortage can cause supply disruptions in the near future[5], while synthetic graphite is neither scalable nor sustainable. The ideal anode material would be metallic Li, which exhibits the highest gravimetric capacity (3860 mAh/g), and the lowest electrochemical potential of -3.04 V vs. standard hydrogen electrode[6] among all anodes. However, metallic Li suffers from poor cycle stability due to extensive side reactions with the electrolyte[7], as well as dendrite formation and growth causing short circuits leading to battery failure[8].

A promising alternative is to use lithiophilic metal substrates that reduce the metallic Li nucleation energy[9–12], while providing high bulk and surface Li diffusivities[9,10,13]. These substrates are reported to act as sinks for Li due to their high solid solubility[11], while homogenizing the Li-ion distribution[14], thereby avoiding dendrite formation. Elements with high solid solubility in body centered cubic lithium (BCC-Li), such as Mg, Al and Ag[12] could replace or simply coat current collectors.

The use of Ag-based thin layers, usually as a composite with carbon, significantly improves the energy density and the cycle stability of anode-less Li-metal batteries i.e., the Li anode is plated during lithiation[15]. Despite Li plating/stripping being the main process underpinning the performance of these batteries, the lithium ingress into the substrate and Li removal from the substrate, respectively, occurs simultaneously[11]. In most materials, lattice defects control many materials' properties, particularly elementary atomistic transport [16–18]. Previous attempts to understand the effect of the substrate's microstructure, i.e. the hierarchy of material grains and defects, on Li plating/stripping[19] and (de)alloying of the Li-solid solution[20], focused on second phases formation. Insofar, grain boundaries had only been hypothesized to act as fast Li diffusion paths during lithiation[21]. Previous studies could not directly resolve the spatial distribution of Li, and experimental results were interpreted based on equilibrium thermodynamics to understand phase transformations during (de)lithiation[10,22].

Here, we introduce a novel workflow that combines Li-Ag micro-diffusion couples made directly in-situ in a scanning electron microscope–focused ion beam instrument (SEM-FIB) with (scanning) transmission electron microscopy ((S)TEM) and atom probe tomography (APT) characterization. APT measurements provide 3D compositional maps with sub-nanometer resolution[23,24] and the ability to discern up to 80% of all the ions a material consists of according to their mass-to-charge, also including light elements such as Li[23,25], whereas crystallographic and structural information are provided by (S)TEM[26,27]. In addition, electron energy loss spectroscopy (EELS) was performed to visualize the Li distribution on a larger scale. Our Li/Ag model diffusion couples allows for the quantification of Li at grain boundaries to understand their role on lithiation, while avoiding secondary factors, such as the nature of the electrolyte, and environmental contamination[28–31]. Our results demonstrate that the lithiation is kinetically-controlled by grain boundary diffusion, with confined formation of Li-rich regions, containing up to >93.8 at.% and does not follow bulk equilibrium thermodynamics. Lithiation and phase transformations in the grain interior are kinetically and possibly chemo-mechanically hindered, implying that the earlier reported electrochemically measured high Li-diffusivities in Ag must result from the intense grain boundary or surface lithiation mechanism observed in this work. Random high angle grain boundaries are favorable sites for the nucleation of Li-rich regions, and hence controlling the grain size and grain boundary structure emerges as an important design criterion for next generation substrates for anode-less batteries.

**Results**

**Diffusion couple**

A fine-grained (average 1.5 µm) pure Ag (> 99 at.%) and a <100> single-crystal specimens taken from annealed (600°C/5 h) Ag thin film were used as substrates for lithiation (Supplementary Figure 1). The detailed experimental procedure can be found in the supplementary information. In short, Ag was lifted out and welded to a support in the FIB, and was then cut into micro-pillar shaped specimens. Following surface cleaning, a Li-lamella was attached on these substrates by using redeposition welding at -190°C[32], forming micro-diffusion couples at the contact, as illustrated in Figure 1a. The stage was heated to 30°C to trigger lithiation of the silver substrate. After 1 h, the fine-grained Ag pillar non-uniformly swells at certain grain boundaries (Figure 1b, right). The <100> Ag single crystal reacted with Li until 4 h without any volume expansion indicating alloying, Figure 1c. Finally, each Ag pillar is cooled down to -190°C. APT specimens were finalized, as displayed in Figure 1d, so as to have their tip in selected sections of each pillar to monitor different stages of the reaction. A micro-diffusion couple with fine-grained Ag was made on a TEM lamella and subjected to lithiation for 40 min at 30°C, leading to a volume expansion of 40–80%, Figure 1e, followed by cooling to -190°C before thinning. Supplementary Figures 2–4 are snapshots of the lithiation process into specimens for APT and TEM.

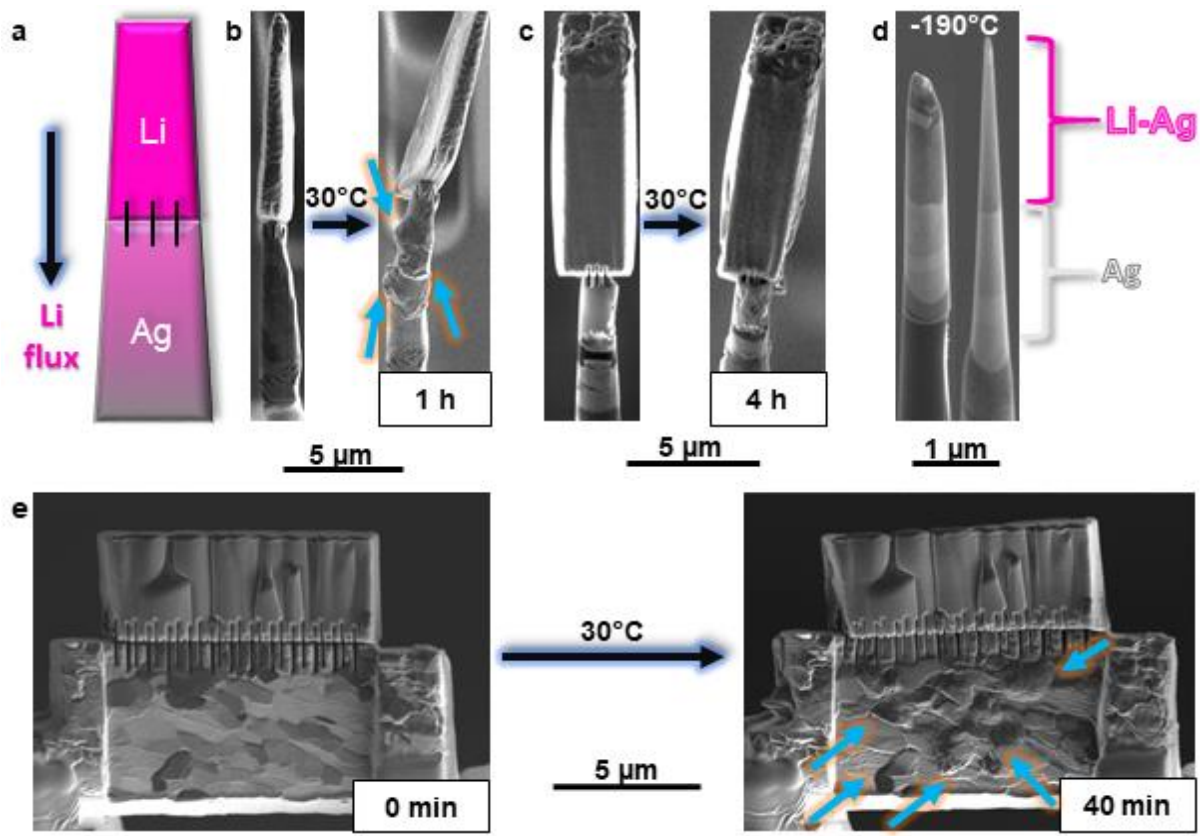

**Figure 1 –** (a) Ag micro-pillar mounted on Si posts used for APT characterization. (b) Micro diffusion couple of Li in contact with Ag at 30°C. Three stripes are milled at cryogenic temperature (-190°C) to ensure connection between both metals. After reacting for 1h, the Ag pillar locally swells by 40-80 % in volume near grain boundaries (blue arrows) due to the spontaneous lithiation. (c) Micro diffusion couple with <100> Ag as substrate. After 4 h of reaction, no swelling but surface roughening is observed. (d) Sharpening into APT specimens, with Li-rich regions in dark contrast, and final APT

specimen used for characterization. (e) Diffusion couple on a TEM lamella of fine-grained Ag, attached with 24 stripes to ensure connection. After 40 min, heterogeneous swelling is observed due to Li alloying.

**Chemical composition evolution**

After 1 h of reaction at 30°C, as shown in Figure 1 and Supplementary Figures 2-3, the swelling of the Ag pillars is concentrated at the Ag grain boundaries, with local volume expansion typically ranging between 40–80% (Supplementary Table 1). The darker contrast in the electron micrograph in Figure 2a indicates Li-enriched regions at grain boundaries. At approximately 2 µm below the diffusion couple interface, the APT analysis in Figure 2b evidences a heterogeneous Li distribution within a single grain boundary. Composition profiles through a precipitate, presented in Figure 2c, show that the Li content reaches values up to 87.6 at.% (orange arrow in Figure 2b), while the surrounding Li-rich grain boundary in Figure 2d shows a decoration value near 32.5 at.% Li (black arrow in Figure 2b). In a zone 4 µm further below, another APT dataset also shows Li enrichment in a GB, Figure 2e. The Li content in this region is approximately 25.8 at.%, Figure 2f, and Li-rich precipitates contain near 58.6 at.% Li, Figure 2g. Regardless of the distance to the diffusion couple, the Li-rich precipitates are nearly spherical in shape.

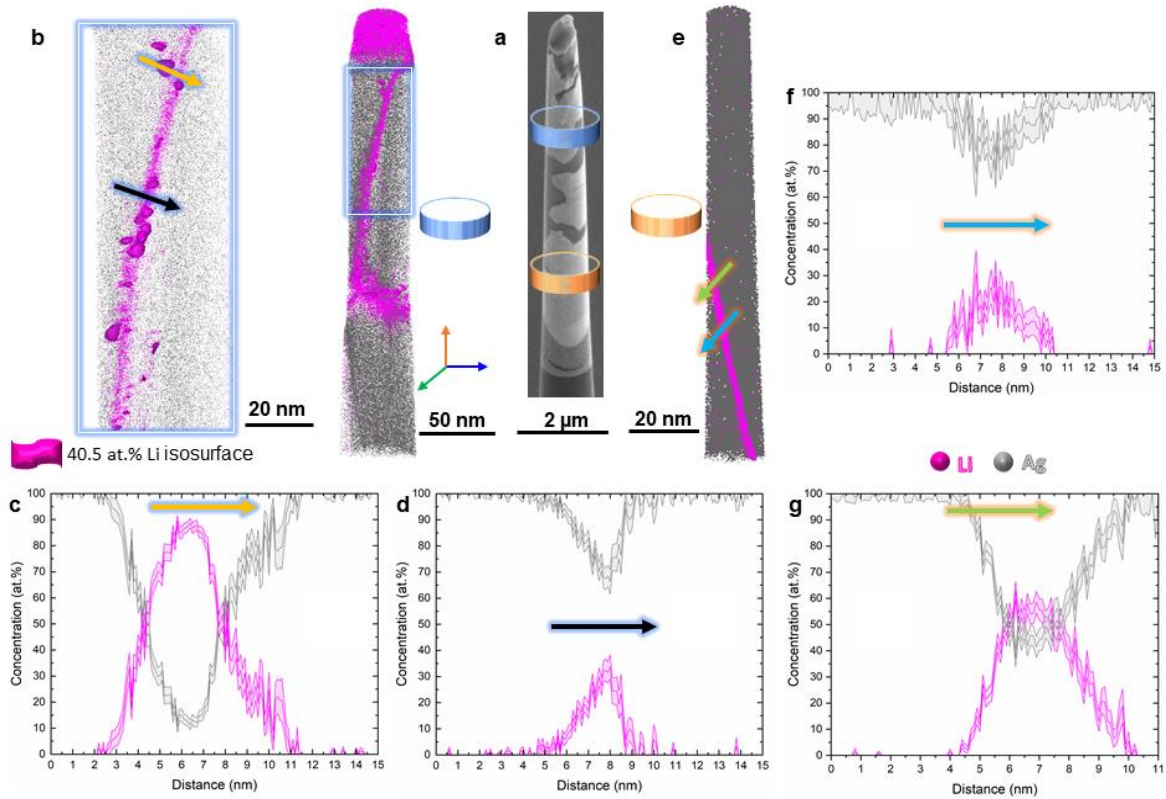

**Figure 2 –** APT results of Li-rich grain boundaries. (a) Orange and blue rings indicate where APT measurements were performed in a typical micro-diffusion couple. (b) Top region, where three grains are visible. Some regions reach up to (c) 87.6 at.% Li, while other areas along the grain boundary exhibit lower amount (down to 20 at.% Li), as indicated by the 1D concentration profile along the orange arrow. (d) 1D chemical composition profile in another region, from the same grain boundary as (b), with more homogeneous Li distribution reaching 32.5 at.% Li in the black arrow from (b). (e) A grain boundary nearly 6 µm below the diffusion couple is visible from the high local Li concentration. 1D composition profiles in (f) through a precipitate-free region of the same grain boundary, with up to 25.8 at.% Li (blue arrow in (e)) and (g) through a Li-rich precipitate containing up to 58.6 at.% Li (green arrow in (e)).

In other regions, such as shown in Figure 3b, the Li content reaches values of even up to 93.8 at.%, indicating possible formation of a BCC-Li nucleus on the grain boundary. The interface is sharp between the Li-rich region and the adjacent FCC-Ag bulk grain interior, which contains only below 0.06 at.%Li. After 14 h of reaction, Li-rich regions formed along the grain boundaries imaged in Figure 3c contain 35–40 at.% Li, as quantified in the composition profile in Figure 3d. Nearly spherical precipitates are also distributed within these bulkier Li-rich phases, and contain up to 88.2 at% Li, Figure 3d, while a remaining unreacted matrix still persists.

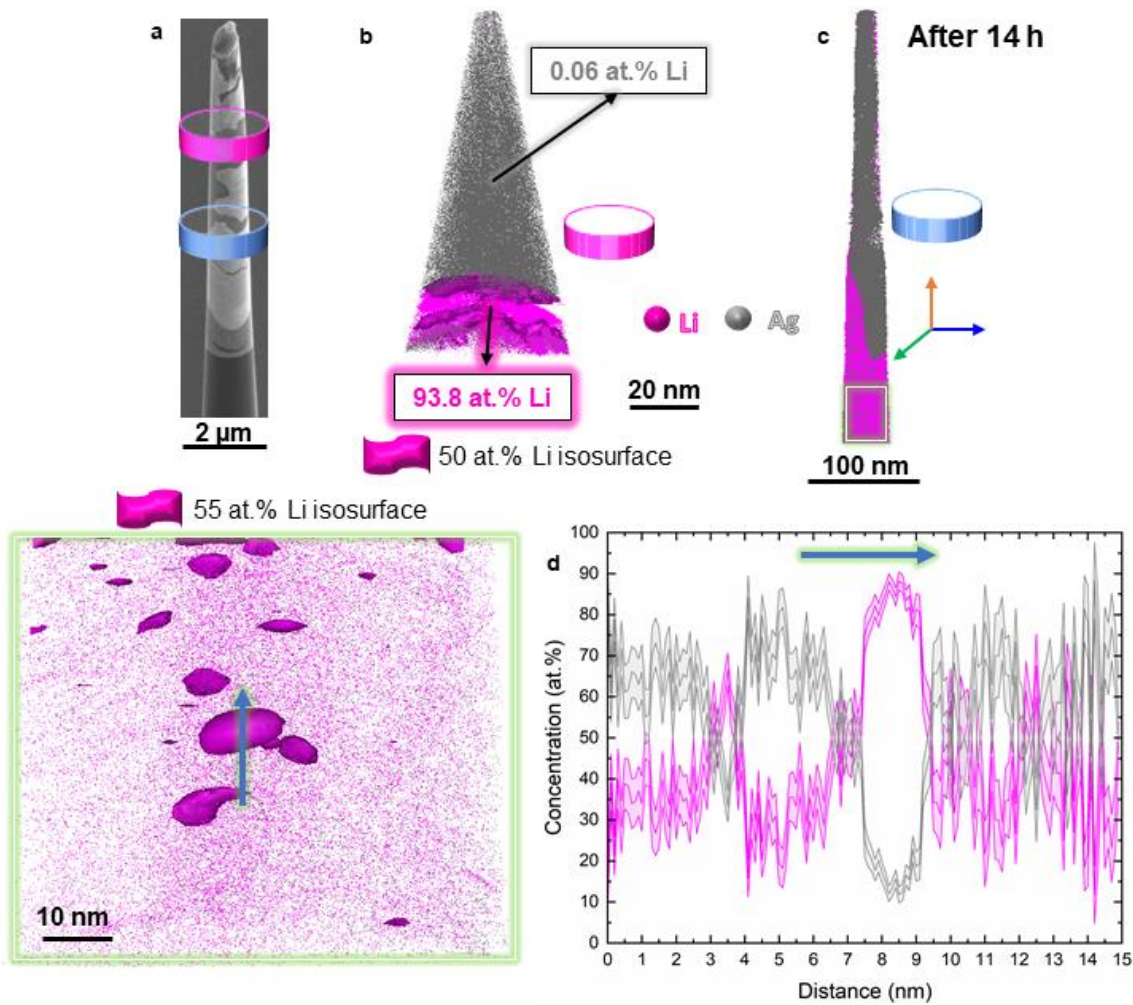

**Figure 3 –** (a) The approximate regions from where APT measurements were performed are shown by the blue and magenta rings in a typical micro-diffusion couple. (b) A second region exhibits a Li-rich phase nucleated at a former Ag grain boundary reaching up to 93.8 at.% Li while the surrounding Ag grains remain non-lithiated. (c) APT data reconstructions from an interface between a lithiated and non-lithiated regions after 14 h reaction. The lithiated region is magnified in the green box (c), with spherical precipitates highlighted by the 55 at.% Li isosurface. (d) A region of interest along one of the spherical precipitates shows the chemical composition reaching up to 88.2 at.% Li in a 1D chemical profile (blue arrow).

**Microstructural evolution**

In the scanning electron microscope, Figures. 4a–b, we used a combination of transmission Kikuchi diffraction (TKD) and X-ray energy-dispersive spectroscopy (EDS) on a diffusion couple created on the TEM lamella shown in Figure 1e. The darker regions in the backscattered electrons image, enriched in Li, percolate along the grain boundaries, leaving numerous grain interiors non-lithiated even though they were close to the diffusion couple interface. There is also a highly lithiated region near the bottom of the sample. Figure 4c is the superimposition of the colored out-of-plane inverse pole figure map obtained through crystallographic analysis of the FCC-Ag TKD patterns and an Ag EDS map in grayscale. In correlation, these maps reveal both, the structure and the chemical features with local resolution. Complementary selected-area electron diffraction (SAED) lattice structure analysis conducted in the TEM, in Supplementary Figure 5, supports the rationale that the lithiated region is mostly the FCC-Ag phase, with a low volume fraction of $Ag_3Li$ precipitates usually seen agglomerated together (Supplementary Figure 5b). The lithiation process refines the grains (372 ± 202 nm) compared to the non-lithiated areas (1465 ± 581 nm) as readily visible in the close-up in Figure 4d. Correlation of the misorientation angle of the grain boundaries with the Ag EDS signal suggests that random high angle grain boundaries are preferentially lithiated; twin and low angle grain boundaries (which consist of dislocation arrays) show negligible tendency for lithiation (Figure 4e).

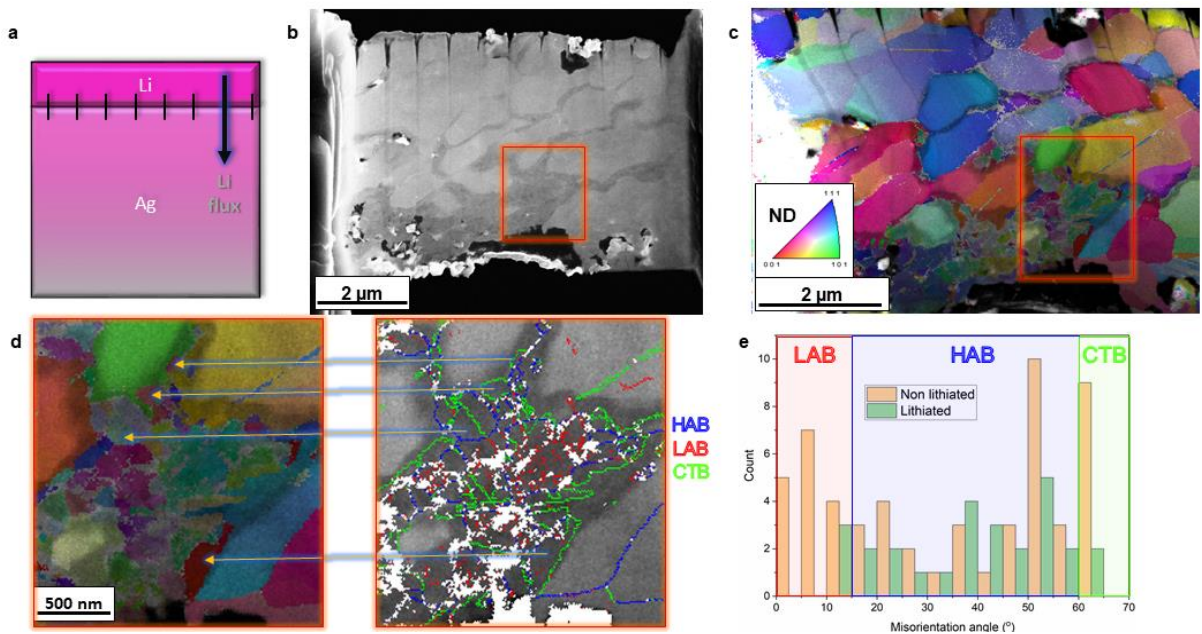

**Figure 4 –** (a) Schematic from the diffusion couple setup and the corresponding position of Li and Ag relative to the TEM/TKD lamella. The Li previously present at the top of the Ag was removed prior to the lamella thinning. (b) Overview of the TEM lamella used for Transmission Kikuchi diffraction. Darker regions correspond to Li-rich areas, which percolate along the grain boundaries. (c) Inverse pole figure color scale (directions perpendicular to the observed surface, or ND – normal direction) with EDS Ag signal gray scale (darker regions indicate lower Ag amount, i.e., Li-rich regions). (d) Zoom-in from the red box in Figure 4b/Figure 4c displaying the TKD map left and the EDS Ag map (signal as gray scale) right with the corresponding grain boundary map overlapping. HAB: high angle boundaries in blue; LAB: low angle boundaries in red; CTB: coherent twin boundaries in green. (e) Relationship between the misorientation angle and the observation of lithiation in a grain boundary. Low angle (Θ ≤ 15°, red) and twin grain boundaries (green) are mostly not lithiated, whereas high angle grain boundaries are mostly lithiated (Θ > 15°).

At higher resolution, high-angle annular dark field (HAADF)-STEM of the lithiated regions near the Ag/Li-Ag interface in Figure 5a shows Ag-rich grains, with a bright contrast, and Li-rich regions appear darker. Figure 5b is a close up on the region-of-interest marked by an orange box in Figure 5a, which contains dark nanometric features, with one indicated by the magenta arrow. The green arrow marks the region where a twin boundary is expected from the grain boundary map in Figure 5c. The contrast in the EELS map indicates a possible compositional shift in the twinned region (Figure 5d). The corresponding EELS spectrum for Li-rich and Ag-rich regions is shown in Supplementary Figure 6. The small Li-rich features are found near the Ag/Li-Ag interface. Unfortunately, their crystal structure could not be resolved due to their small size and low volume fraction and would require higher resolution investigation at lower temperature to minimize damage from the electron beam. Similar Li-rich features observed in the APT data (Figure 2c) reached up to over 88 at.% Li.

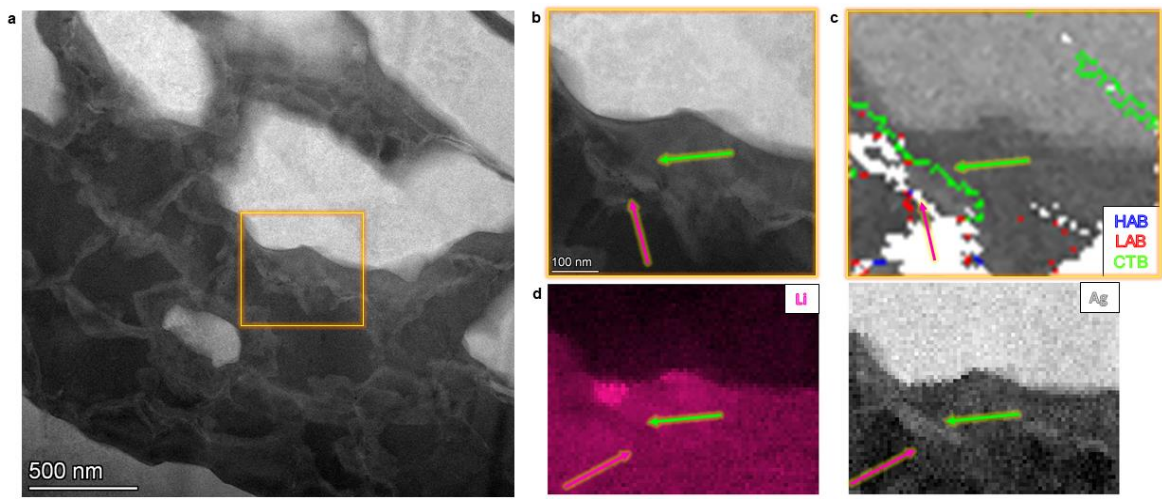

**Figure 5 –** Ag-Ag/Li interface observed by TEM, TKD and EELS. (a) Low magnification and (b) zoom-in HAADF images; (c) Boundary map from TKD. HAB: high angle boundaries in blue; LAB: low angle boundaries in red; CTB: coherent twin boundaries in green. Li-rich regions are visible as nanometric darker precipitates in (b). (d) EELS maps from Li and Ag.

**Discussion**

**Kinetics vs. thermodynamics of grain boundary lithiation**

The Li enrichment at high angle grain boundaries can be predominantly caused by either thermodynamic or kinetic factors. Despite the observation of Li-rich precipitates in the lithiated matrix, we consider only the FCC-Ag phase for the thermodynamic calculations since the grain boundary lithiation in this phase acts as precursor mechanism for the precipitation. We assume that the adsorption isotherm leads to a near solid-solution type of lithiation of the interfaces and, upon reaching a certain local chemical potential, interfacial precipitation can take place.

For the thermodynamic factor, equilibrium grain boundary segregation[33–36] could lead to a local Li-enrichment due to energy minimization of the interface (hence, total) free energy through adsorption of solute atoms[37]. This can be rationalized based on the enthalpy of mixing of the binary system, i.e., the excess enthalpy caused by the interaction between the segregating atoms in regular solutions[37], where a negative value indicates an exothermic reaction. The calculated enthalpy of mixing indicates a thermodynamic tendency to form a homogeneous solid solution (negative enthalpy) for FCC-Ag across the entire Ag-Li composition range (Supplementary Figure 7a), which is reflected in the wide FCC-phase range in Supplementary Figure 7b.

This prediction is in contradiction with our experimental results in Figures. 1-4. Consequently, the observed preferential lithiation along grain boundaries cannot be elucidated through principles of bulk or respectively interfacial equilibrium thermodynamics in the solute limit alone. Rather, the quite extreme decoration (with a Li concentration of up to 93.8 at.% Li adjacent to unreacted Ag) and phase formation phenomena observed here seem to be dominated by kinetic (non-equilibrium grain boundary segregation), i.e., enhanced grain boundary diffusivity, and possibly chemo-mechanical factors. This is extremely likely to also be the case for lithiation with other electrode and electrolytes employed in batteries, especially when operated at higher current densities, that could further increase the Li flux and reinforce the importance of kinetics.

**Role of grain boundary on kinetics - estimation of Li diffusivity in Ag**

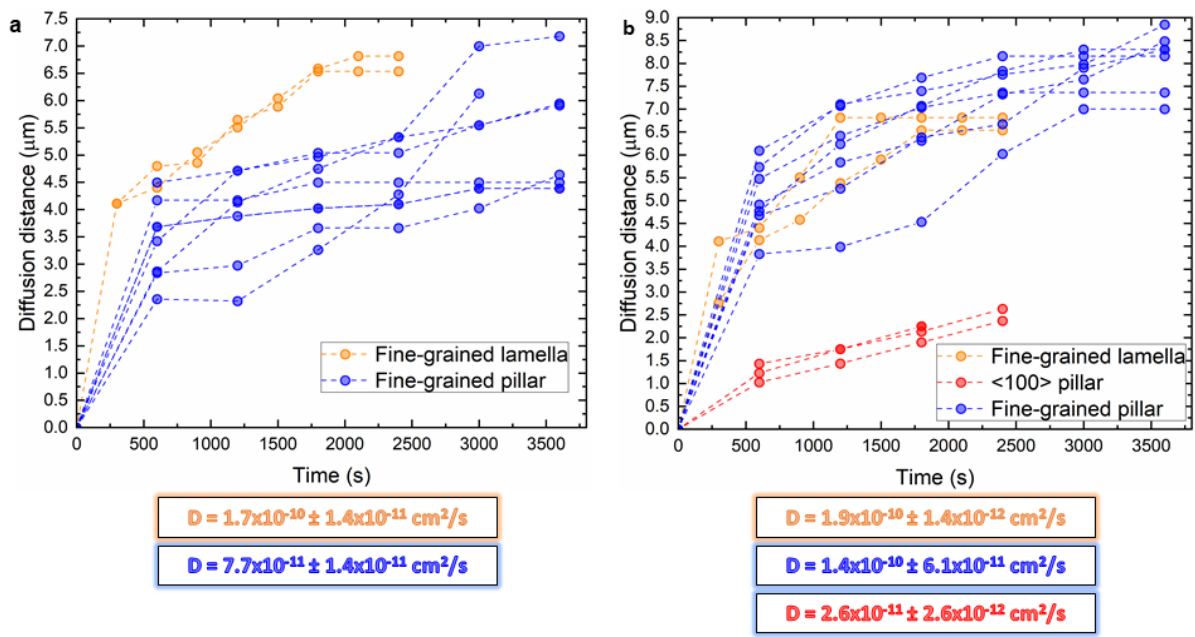

**Figure 6 –** Li diffusion distance considering (a) only Li alloying, i.e., the farthest swollen region observed. (b) Li diffusion on the surface, where the propagation length was considered as the farthest region exhibiting some swelling or contrast change on the surface. <100> and fine-grained pillars are considered, as well as fine-grained lamella. Data from <100> single crystals are only available in (b) since they did not exhibit any swelling caused by Li alloying. The diffusion distance can become constant when it propagated along the whole length in (b) or if only a certain grain boundary was lithiated in (a). An example on the estimation of the diffusion distance is given in Supplementary Figure 8.

The effect of grain boundaries on the Li diffusivity was quantified by the diffusion distance observed in the micro-diffusion couple. Two scenarios are considered. First, we recorded the progression of the localized swelling associated to alloying, as plotted in Figure 6a, ignoring the surface diffusion. Second, we considered contrast changes associated to the surface diffusion of Li. This was achieved by imaging a single location by using the ion beam as exemplified in Supplementary Figure 8. The resulting evolution of the distance vs. time is plotted in Figure 6b. Given the parabolic shape of the L vs. t plot (L being the propagation length, t the time), the grain boundary/free surface lithiation of Ag is a diffusion-controlled reaction[30,31,38]. For such a reaction, the diffusivity can be calculated as the slope of $L^2$ vs t plot or as $L^2/4t$[38]. We employed mainly the former scenario, except in the case where a single grain boundary was lithiated in Figure 6a, i.e., when the diffusion distance is constant after some time.

The <100> single crystal does not exhibit Li alloying even after exposure times of up to 4 h of reaction, as confirmed by APT measurements (Supplementary Figure 9); yet, Li still diffuses along the surface of the pillar, and we estimate a free-surface diffusivity of D = $2.6 \times 10^{-11} \pm 2.6 \times 10^{-12}$ cm$^2$/s. We also find that surface diffusion (Figure 6a) propagates farther than the alloyed regions, by a factor of about 1.1–2 depending on the substrate's shape (Figure 6b). Pillars often contain only a single grain boundary that gets alloyed by the micro-diffusion couple, and the constant diffusion distance in Figure 6b is due to the absence of a grain boundary network along which Li could diffuse farther, as exemplified in Figure 4.

Smaller cross sections, i.e. relatively larger surface/volume ratios, favor faster diffusivity due to the predominant surface diffusion compared to the bulk[28,30,38]. Yet the lamellar diffusion couple, Figure 1e, exhibits a longer diffusion distance, which can be rationalized by the presence of a grain boundary network, offering a fast diffusion path for Li. The similarity between diffusion coefficients for lamellas when comparing only alloying, D = $1.7 \times 10^{-10} \pm 1.4 \times 10^{-11}$ cm$^2$/s, with the surface diffusion of <100> pillar, D

= 2.6x10$^{-11}$ ± 2.6x10$^{-12}$ cm$^2$/s, is another evidence that diffusion through a grain boundary network can be as fast as free-surface diffusion[39].

**Role of lithiation-driven grain-scale mechanics – kinematics, plastic deformation and strain hardening**

The increased Li content in the Ag anode as the lithiation proceeds has two major consequences on the microstructure. First, it creates a strain gradient between the Li-rich and Li-poor regions, as the crystalline FCC structure locally expands by 40–80 vol.% to accommodate the inserted Li. This massive volume increase in the chemically highly decorated regions along the grain boundaries causes a kinematic displacement perpendicular to the interface planes, as schematically depicted in Figures 7a–c. Such kinematic displacement acting on the adjacent grains is different from the kinematic states observed when the whole bulk grain is lithiated (see the grain swelling in Figure 7d), which in turn creates a less intense local stress state in the adjacent, unreacted Ag grains.

The strain gradient at lithiated grain boundaries is accommodated by arrays of geometrically necessary dislocations (GND). These defect patterns can be quantified in terms of the kernel average misorientation (KAM) values obtained from TKD. The KAM value is a metric for local orientation gradients which result from the volume mismatch, as displayed in Figure 8a. The average GND density in the Ag grains adjacent to the Li-rich areas at GBs is 7.6 x 10$^{14}$ m$^{-2}$ (see Supplementary Information), comparable to severely deformed Ag[40]. Extra statistically stored dislocations are also nucleated, although they do not contribute to the accommodation of strain gradients[41–43].

The unreacted Ag grains prevail even after 14 h Li exposure (Figure 3c) and the high dislocation density in regions adjacent to the massively lithiated grain boundary areas indicates that a mechanical effect (elastic stresses) may aid in hindering the reaction within the grain interior. Elastic stresses are caused by misfit, such as differences in lattice parameter, heterogeneity in plastic deformation, differences in thermal contraction/expansion or strain caused by phase transformation[44]. After the grain boundaries are favorably lithiated due to the enhanced kinetics following a diffusion-controlled reaction (Figure 6), the reaction switches to an interface-controlled mode for the lithiation in the grain interior. This transition in the reaction mode occurs because of the strain hardening that results from the intense plastic activity in the adjacent Ag grain, as indicated by the high dislocation density (Figure 7c).

A complex triaxial stress state happens in both, the lithiated and unreacted regions, due to the elastoplastic behavior and inelastic stress relaxation by dislocation nucleation and slip (Figure 8). The von Mises stress is a criterion for plastic strain onset, which converts a triaxial stress state into a single stress value to be compared with the yield stress of Ag under uniaxial stress conditions observed in a standard tensile test. A local higher von Mises stress suggests likewise higher residual elastic stresses[45].

The "flat" interfaces highlighted by dashed arrows in Figure 8 indicate an elastic stress peak within the unreacted Ag, near the reaction interface, with components of compressive normal stress and shear (triaxial state). In contrast, the interfaces near the grain corners, crossed by solid arrows, exhibits a triaxial stress concentration within the lithiated region. In both cases, the shear stresses are caused mostly by the elastoplastic behavior concentrated at the interface. At the corners, the elastic stresses are caused mainly by geometrically necessary dislocation arrays within the lithiated region, while the compressive and shear stresses within the unreacted Ag near the flat interfaces reduces the driving force for further lithiation.

The stress build-up is confirmed by both the superior GND density (Supplementary Figure 10a) near the reaction interface, as well as the local higher von Mises stresses (Figure 8), which gradually decrease towards the unreacted Ag grains. The stresses caused by the expanded lithiated regions may be below the yield stress of the remaining hardened Ag grains and cannot be released by plastic deformation. As a result, the Ag grains are only in the purely elastic regime, resulting in the increased compressive stress in the adjacent lithiated region, thus mechanically decreasing the driving force for further lithiation[46,47].

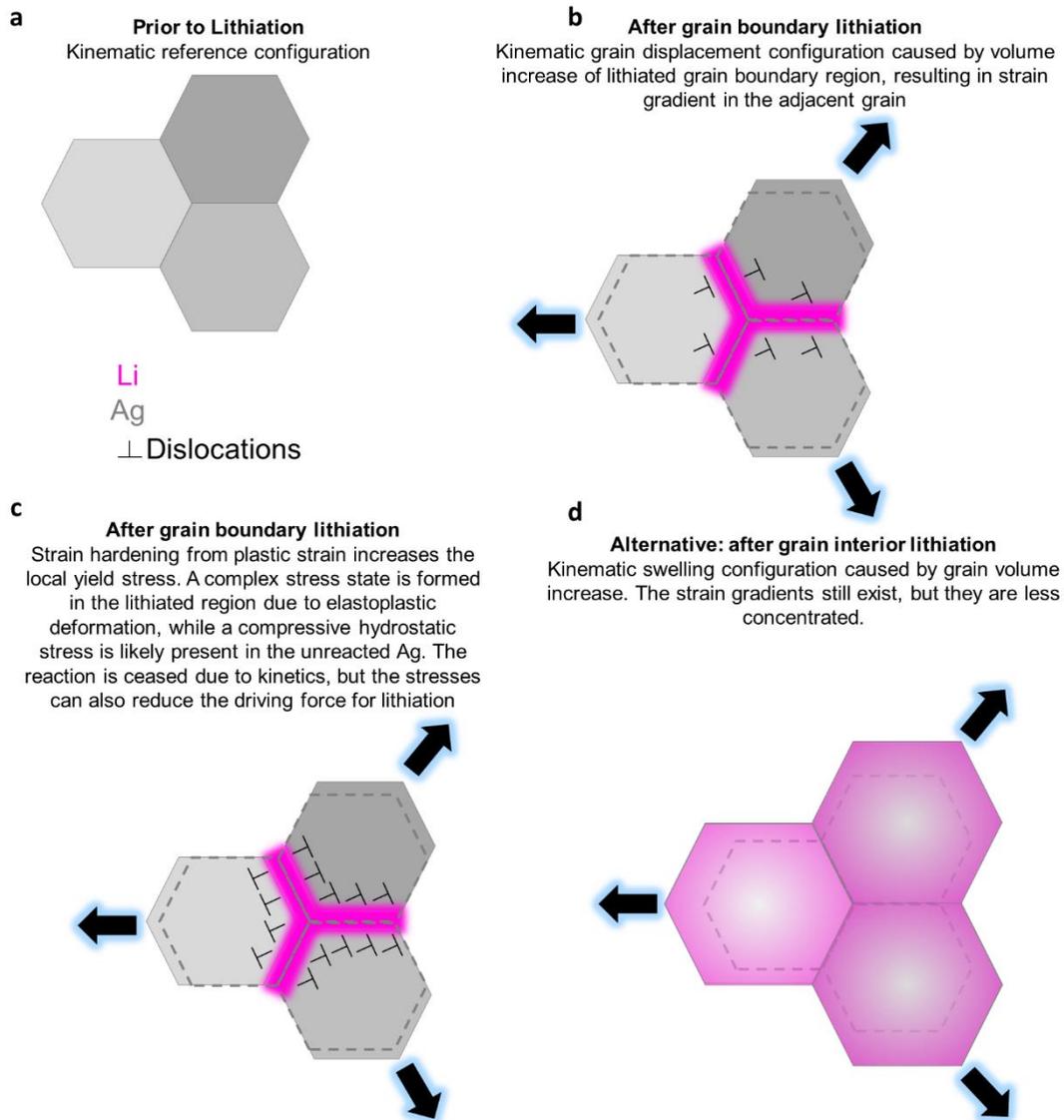

**Figure 7 –** Schematics of the difference in grain kinematics depending on the lithiation mode. (a) Reference configuration prior to lithiation. (b) After grain boundary lithiation, a kinematic grain displacement takes place, caused by volume increase of lithiated grain boundaries, resulting in plastic strain in the adjacent unreacted grains. (c) After a certain point of grain boundary lithiation, the plastic strain causes enough strain hardening to increase the yield stress of the adjacent Ag grains. This increases the hydrostatic stresses in the Ag grain and the triaxial stresses in the adjacent lithiated regions, reducing the driving force for further lithiation and ceasing the reaction. (d) Alternatively, the kinematic swelling of grains would take place if the lithiation was taken mainly in the grain interior.

Second, the lithiation increases the homologous temperature, i.e. the ratio between the experimental temperature condition (30°C) and the actual melting point for the specific local chemical composition, as seen in Supplementary Figure 7b. All diffusion barriers scale with the local (concentration-dependent) melting point, such as the vacancy formation and diffusion energy. At increased homologous temperatures, the increased diffusivity and vacancy generation activates microstructure restoration mechanisms to decrease the defect density, i.e. dislocation recovery[48,49] and recrystallization[50,51].

While recovery reduces the defect density mainly through dislocation annihilation and rearrangement, recrystallization is based on the nucleation and expansion of "defect-free" nuclei mediated by a moving high angle boundary[52]. In FCC-materials with medium-to-low stacking fault energy, twin boundaries are formed during recrystallization[53,54]. As observed in Supplementary Figure 10b, many grains with low grain average misorientation are formed within lithiated regions and exhibit a twin relationship with one of the adjacent grains. Meanwhile, they are separated from other grains with a high average misorientation by high angle boundaries. These features are characteristic from the growth accident model for twin boundary formation[55], where a twin boundary is left behind a moving (high angle) grain boundary[56]. The increased Li content reduces the stacking fault energy of the matrix lithiated region[57] thereby facilitating the formation of such twin boundaries during recrystallization[58].

The gradually increasing plastic strain and homologous temperature with the progressive Li alloying in the lithiated area triggers dynamic recrystallization[59]. The dynamic nature is caused by the concomitant occurrence of recrystallization with the continuous plastic deformation, where the fresh "strain-free" grains are deformed as soon as they are formed[60].

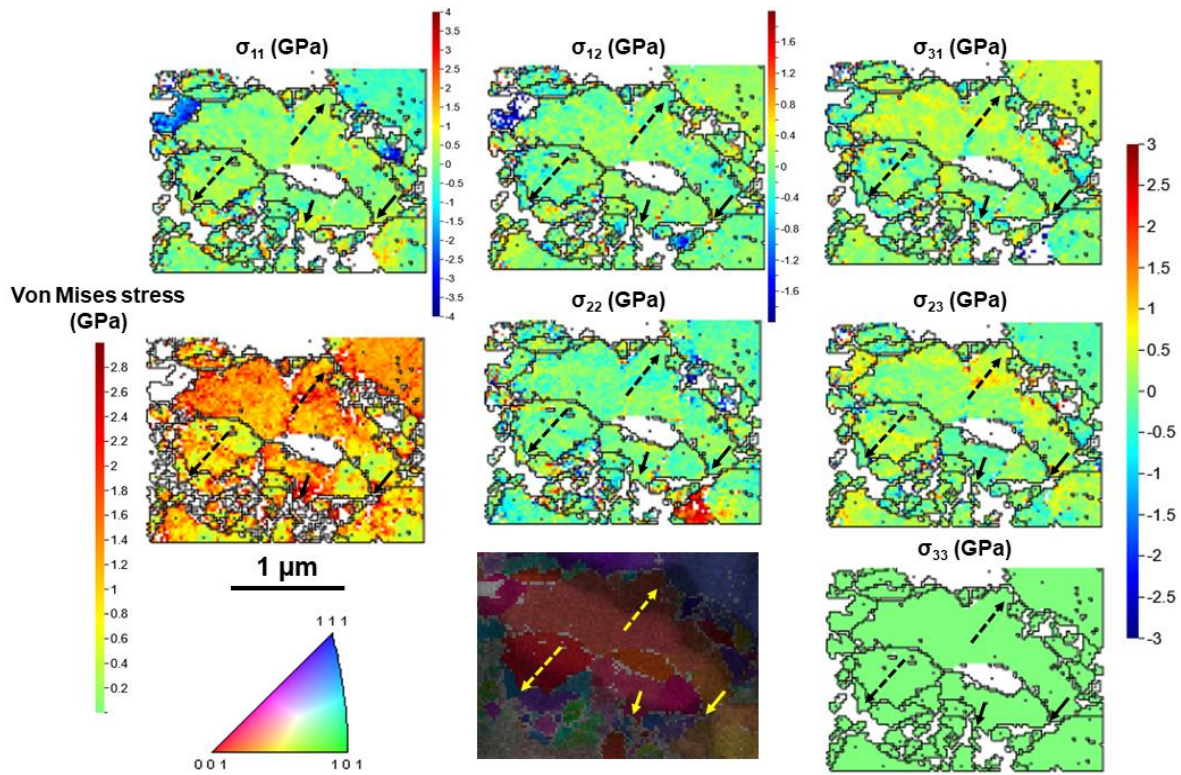

**Figure 8** – Local elastic stresses calculated from cross correlation, indicating higher von Mises stresses near the interface between lithiated and unreacted Ag grains, as indicated by arrows crossing such interfaces in all images. A complex stress state is caused by the volume expansion of the lithiated regions and the elastoplastic behavior of both lithiated and unreacted regions. Compressive stresses are negative, while positive stresses are tensile.

**Generalization in the context of a battery**

Despite the advantage of tracking the Li ingress at specific lattice defects with no contamination or under the influence of secondary factors, the use of diffusion couples has limitations when compared to a battery in operation. First, no electrolyte is directly used and any effect from the solid electrolyte interphase and electrolytes are not reproduced in our experiment, despite their strong influence on the electrochemical behavior[61].

Second, the absence of an applied potential in the diffusion couple changes the phase transformation pathway, both thermodynamically and kinetically, since the reaction is purely chemical in a diffusion couple. In the diffusion couples, we only observed FCC-Ag and some minor $Ag_3Li$ phases (Supplementary Figure 5). The nucleation of Li-rich particles or phases apparently follows this reaction sequence: FCC Ag-Li → FCC Ag- (20-40 at.%)Li → unknown phase with up to 93.8 at.% Li. The transformation sequence matches with the predicted stable phases at room temperature in a revisited Ag-Li phase diagram[62]. Our experimental results indicate that over-saturation[63] of the FCC phase does not take place. Instead, the spherical shape of Li-rich precipitates within the lithiated boundaries indicates that BCC phase should nucleate by a classical two-phase mechanism. Despite the strong kinetic and mechanical effect during the overall lithiation, the local phase transformation pathway within a single grain boundary appears to follow a near-equilibrium condition since it is kinetically allowed.

Applying an artificial potential during lithiation lowers the potential towards more negative values (or increases the chemical potential of Li), driving the formation of Li-Ag phases with increased Li content depending on the applied current density[11,22,64]. At some point, the potential would be negative vs. Li/Li$^+$ and metallic Li would be plated in the Ag, as commonly observed in Ag[10,11] and Ag-C[15,65] interlayers/substrates. Due to the applied potential, BCC-Li nucleation within the Ag substrate would also be intensified.

The diffusion at grain boundaries should follow the A-C classification from Mishin[66], where in the C regime the diffusion is constrained to the grain boundary plane. After prolonged times, the Li diffuses towards the grain interior, reaching the adjacent regions from the grain boundary (B) and, finally, consuming the whole grain (A). This sequence causes the kinematic grain displacement (Figures 7b-c), resulting in confined local plastic strain at the Ag grains adjacent to the lithiated boundaries, possibly aiding at ceasing the reaction in regime C after a certain elastic stress is achieved. With an applied potential, we expect that grain boundaries will still be preferentially lithiated at first given kinetic and likely chemomechanical reasons, a mechanism that intensifies at higher current densities.

As observed in Figure 4e, not all grain boundaries are effective to host Li and later nucleate Li-rich regions. Only high angle grain boundaries (Θ ≥ 15°) are effectively preferentially lithiated due to their higher energy and higher diffusivity. Low angle boundaries (2 < Θ < 15°) and coherent twin boundaries, characterized by lower boundary energy[67] and lower diffusivity[68], are not lithiated. This is detrimental for lithiation and Li transport due to two factors. First, a lower boundary energy is detrimental for the ability of a grain boundary to reduce the nucleation overpotential[69] by heterogeneous nucleation. Second, the lower grain boundary energy causes a lower tendency for grain boundary enrichment[70], and hence lower grain boundary diffusivity of Li, hindering the Li transport within the substrate. Designing a nanocrystalline substrate with the highest amount of random high angle boundaries possible would be a good option for superior kinetics at a given current density.

When considering the substrate/anode design, Ag is often used in the form of nanoparticles or in thin film/foil form. A composite electrode offers free volume for Ag nanoparticle expansion upon lithiation, reducing stress evolution in the nanoparticle. Contrastingly, thin films or respectively foils will develop higher stresses due to the absence of free volume that could compensate for the gradual concentration-dependent expansion of the Ag crystals, resulting in intense strain hardening and, consequently, high elastic stresses, associated also with plastic relaxation processes, i.e. nucleation and accumulation of dislocations. The lithiation of grain interiors would be more difficult and crack formation could take place to alleviate such elastic stresses, possibly also aided by local stress peaks that may build up caused by the presence of arrays of geometrically necessary dislocations as described above. Alternatively, a limited lithiation at grain boundaries could take place, possibly compromising the performance of thin film/foil.

In summary, we used micro-scale Li-Ag diffusion couples to elucidate the critical role of grain boundaries in the local enrichment of Li on Ag substrates. Our findings reveal that FCC Ag-Li solid solutions predominantly form at random high-angle grain boundaries ($\Theta \geq 15°$), with Li concentration up to 40 at.%. Such Li-enriched boundaries host spherical precipitates reaching up to approximately 88.2 at.% Li, while some FCC-Ag grain boundaries exhibit confined phases reaching >93.8 at% Li region. Low angle boundaries ($2° < \Theta < 15°$) and coherent twin boundaries are not lithiated due to their lower segregation energy and Li diffusivity. While Li did not show any alloying within a <100> Ag single crystal, the combination of grain boundary and surface diffusions yield a superior diffusivity compared to surface diffusion alone. Thus, grain boundary character and density could be used to design optimized substrates/anodes for enhanced lithiation kinetics.

The lithiation takes place mainly along the grain boundaries, without lithiation in most of the grain interiors. This effect leads to a corresponding kinematic grain displacement, indicating the predominance of a kinematic (position shifting) effect that translates to corresponding kinetics of dislocation populations which in turn mechanically reduces the driving force upon lithiation. It is unlikely that equilibrium thermodynamics alone can be used to predict the chemical evolution for any electrode/solid electrolyte material employed in a battery. Thus, to fully account the lithiation behavior, both factors should be considered.

**Acknowledgements**

The authors thank Uwe Tezins, Christian Bross and Andreas Sturm for their support at the FIB and APT facilities at MPIE. Authors would like to thank Katja Angenendt for the support in the TKD measurement. We also appreciate Alisson Kwiatkowski da Silva for fruitful discussions. P. Yadav acknowledges Alexander von Humboldt (AvH) Foundation for her postdoctoral fellowship. C. Jung acknowledges the Global Joint Research Program funded by Pukyong National University (202411890001). S.-H. Kim is grateful for supports from KIAT grant funded by the Korea Government (MOTIE, HDR-P0023676).

**Supplementary Material**

**Materials and methods**

**Diffusion couple**

A fine-grained (average 1.5 µm) pure Ag (> 99 at.%) and an annealed (600°C/5 h) Ag thin film were used as anode for the lithiation. Ag thin films were deposited on the surface of Si wafers, which had a 1000 nm thick wet oxide layer of $SiO_2$. Films were deposited by magnetron sputtering from a high-purity Ag target (99.99%) in a radio frequency (RF) gun within a physical vapor deposition (PVD) system (BESTEC, Berlin, Germany). The deposition was conducted at room temperature, achieving a film thickness of approximately 1000 nm. The deposition rate was maintained at 0.6 nm/s. Before the sputtering, the chamber was evacuated to a base pressure of $1 \times 10^{-6}$ Pa. Argon gas was then introduced as the sputtering medium at a flow rate of 40 standard cubic centimeters per minute (SCCM), establishing a working pressure of 0.3 Pa. From the annealed Ag thin film, a 20 µm <100> grain was lifted-out to produce <100> single-crystal specimens (Supplementary Figure 1).

A Xe-plasma focused ion beam (Thermo-Fisher Helios PFIB) was used to lift-out and mount the Ag specimen onto a series of commercial Si coupons, following the first steps of the protocol outlined in Ref.[1], as shown in Fig. 1a. This coupon was transferred to a nitrogen glovebox, where a Li piece was cut and added onto a second holder. Both were transferred through a high vacuum suitcase ($<10^{-6}$ mbar)[2], at room temperature, into a ThermoFisher Helios 5 Ga-FIB.

A Li-lamella was lifted-out under cryogenic conditions, using the protocol detailed in Ref.[3]. After cleaning the surface of the top of the Ag pillar and the bottom of the Li lamella at 30 kV/0.23 nA, both metals were contacted, and subsequently four stripes (1.5 µm length) were milled (30 kV/24 pA) vertically at the Li/Ag interface to guarantee proper connection between the metals, Fig. 1b, forming micro-diffusion couples. A batch of six was made in one session. The stage was heated to room temperature and the silver lithiation takes place after approximately 30 min. After 1 h, the fine-grained Ag pillar pronouncedly swells and the Aquilos 2 (Thermo-Fisher) stage is cooled down to -190°C (Fig. 1b, right), while the <100> Ag single crystal was reacted until 4 h (Fig. 1c). The cooling takes approximately 15 min. Each Ag pillar is sharpened into needle-like specimen suitable for APT, as displayed in Fig. 1d.

A similar micro-diffusion couple was created on a lamella for TEM. An Ag lamella with dimensions 10 x 3 x 6 µm$^3$ was mounted onto a Mo grid in the PFIB. The lamella was air transferred to the Ga-FIB. Under cryogenic conditions in the Ga-FIB, Li was lifted-out and connected to the Ag lamella after surface cleaning. Around 24 stripes were vertically milled (1.5 µm length) to ensure proper contact and adhesion of the Li/Ag interface, Fig. 1d, left. The stage was heated to 30°C to activate the reaction, and maintained at this temperature for 40 min. The Li alloying is visible from the swelling of the Ag lamella (Fig. 1e, right). This was followed by cooling to cryogenic temperatures (-190°C) for thinning, which takes approximately 30 min. Supplementary Figures 2–4 are series of snapshots of the diffusion process into APT specimens and the TEM lamella, respectively.

The Li diffusion distance was estimated as the farthest distance from the Li/Ag interface where a microstructure change was observed in SEM. Two cases were considered: first, the case where Li alloying, indicated by swelling, and Li surface diffusion, characterized by contrast changes on the surface of the Ag substrate are accounted. In the second case, only alloying (farthest swollen region) is considered.

The local volume change was calculated considering a cylinder with constant height, where the swelling takes place mainly along the diameter. The diameter from a swollen region was measured before and after reaction. Some inaccuracies can arise from the rotation and bending of the tip caused by the swelling. This can cause the comparison of before/after diameter in slightly different locations. The height is not precisely constant after lithiation, but it cannot be precisely measured due to the aforementioned rotation/bending of the pillar. The obtained values usually range between 40-80% in volume, although two samples exhibit pronounced local swelling, reaching up to 408.8 vol.% in Supplementary Table 1.

**Supplementary Table 1** – Local volume expansion quantification in grain boundary regions upon lithiation of fine-grained Ag pillars.

| Sample | Initial diameter (µm) | Final diameter (µm) | Height (µm) | Initial volume (µm$^3$) | Final volume (µm$^3$) | ΔV (%) |
|---|---|---|---|---|---|---|
| 1 | 1.24 | 1.67 | 1.18 | 1.43 | 2.58 | 81.4 |
| 2 | 1.47 | 2.43 | 1.82 | 3.09 | 8.44 | 173.3 |
| 3 | 1.53 | 1.82 | 1.13 | 2.08 | 2.94 | 41.5 |
| 4 | 1.16 | 1.56 | 1.44 | 1.52 | 2.75 | 80.9 |
| 5 | 0.9 | 2.03 | 2.62 | 1.67 | 8.48 | 408.8 |
| 6 | 1.5 | 2.02 | 1.62 | 2.86 | 5.19 | 81.4 |

**Atom probe tomography**

Atom probe tomography measurements were performed in a local electrode atom probe LEAP 5000XS (Cameca Instrument Inc.) using the pulsing laser mode. The laser energy was 50–60 pJ, at a pulse repetition rate of 200 kHz with a target detection rate of 1 ion per 200 pulses on average, at a base temperature of 50 K. Data reconstruction and processing was performed in CAMECA's APSuite 6.3.

**Transmission electron microscopy**

Transmission electron microscopy and selected area electron diffraction were acquired with an image-corrected Titan Themis microscope (Thermo Fisher Scientific) operated at an acceleration voltage of 300 kV. Scanning transmission electron microscopy (STEM) was performed at 300 kV on a probe-corrected Titan Themis microscope (Thermo Fisher Scientific). For high-angle annular dark field (HAADF)-STEM imaging, inner-outer collection angles of 73 to 200 mrad was used. Electron energy loss spectroscopy (EELS) spectrum imaging was acquired using a Quantum ERS spectrometer (Gatan), and processed by multivariate statistical analysis[4].

**Transmission Kikuchi Diffraction**

Transmission Kikuchi diffraction (TKD) was performed in a Zeiss Merlin field emission SEM. The electron beam acceleration voltage and current were 30 kV and 2 nA, respectively. A 7.3 x 5 µm$^2$ map was acquired, with a step size of 15 nm, with simultaneous X-ray energy-dispersive spectroscopy (EDS) measurement. The acquired data was treated in the EDAX OIM Analysis 8.6 software. Grain confidence index standardization was used as cleanup procedure, followed by removal of any measured pixel with a confidence index lower than 0.1.

The Ag EDS signal count exhibits a bimodal curve that was fitted with two gaussian curves, and their intercept was used to separated the Li-rich from the unreacted Ag grains. These regions were extracted to calculate the kernel average misorientation (KAM) and geometrically necessary dislocation (GND) density. The KAM increases linearly with the number of nearest neighbors considered in the kernel. The GND density is calculated based on the KAM. Since the GND density for a single sample should not vary depending on the user-chosen parameter, we calculate the Θ$_{true}$ based on[5,6]:

$$\theta_{true} = \theta_{noise} + n\theta_{true}$$

Where n is the size of the kernel (n = 1-4) and Θ$_{noise}$ is the intercept in the y-axis of the linear fit. The average GND density was then calculated as[6]:

$$\rho_{GND} = \frac{\theta_{true}}{b.n.a}$$

Where b is the Burgers vector (b = 0.2892 nm for Ag), n is the size of the kernel (n = 3) and a is the measurement step size (a = 15 nm).

For cross correlation, a similar setup was used. A step size of 25 nm was used for a map of 9 x 5.5 µm² with simultaneous EDS measurement. For each pixel, the Kikuchi diffraction pattern was recorded as an image. For cross correlation, the software CrossCourt 4 was used. Only points with CI > 0.1 were employed for the analysis, while only FCC-Ag was detected. The elastic constants for both unreacted and lithiated Ag grains were considered the same, with $C_{11}$ = 13.5 GPa, $C_{12}$ = 11.44 GPa and $C_{44}$ = 8.78 GPa.

**Thermodynamic calculations**

The software FactSage 7.3 was used to calculate the mixing enthalpy as function of Li amount in the FCC phase from the Ag-Li system with the FTLite database. The equilibrium phase diagram was also calculated for this system. The McLean isotherm model [8–10] was used to predict the average grain boundary segregation energy that would lead to the observed grain boundary enrichment, considering that grains contain 99.5 at.% Ag, as obtained from APT, and a constant temperature of 30°C, Supplementary Figure 6a. The grain boundary segregation energy is around -12 to -12.5 kJ/mol for an average grain boundary segregation of 40–45 at.% Li as measured by APT. The calculated negative enthalpy of mixing for the major FCC phase observed in the lithiated region, plotted in Supplementary Figure 6b, indicates a thermodynamic tendency to form a homogeneous FCC Ag-Li solid solution across the entire Ag–Li composition range, i.e. without segregation. This is reflected in the large solid solution range for this phase in the calculated phase diagram, Supplementary Figure 6c.

**Supplementary Figures**

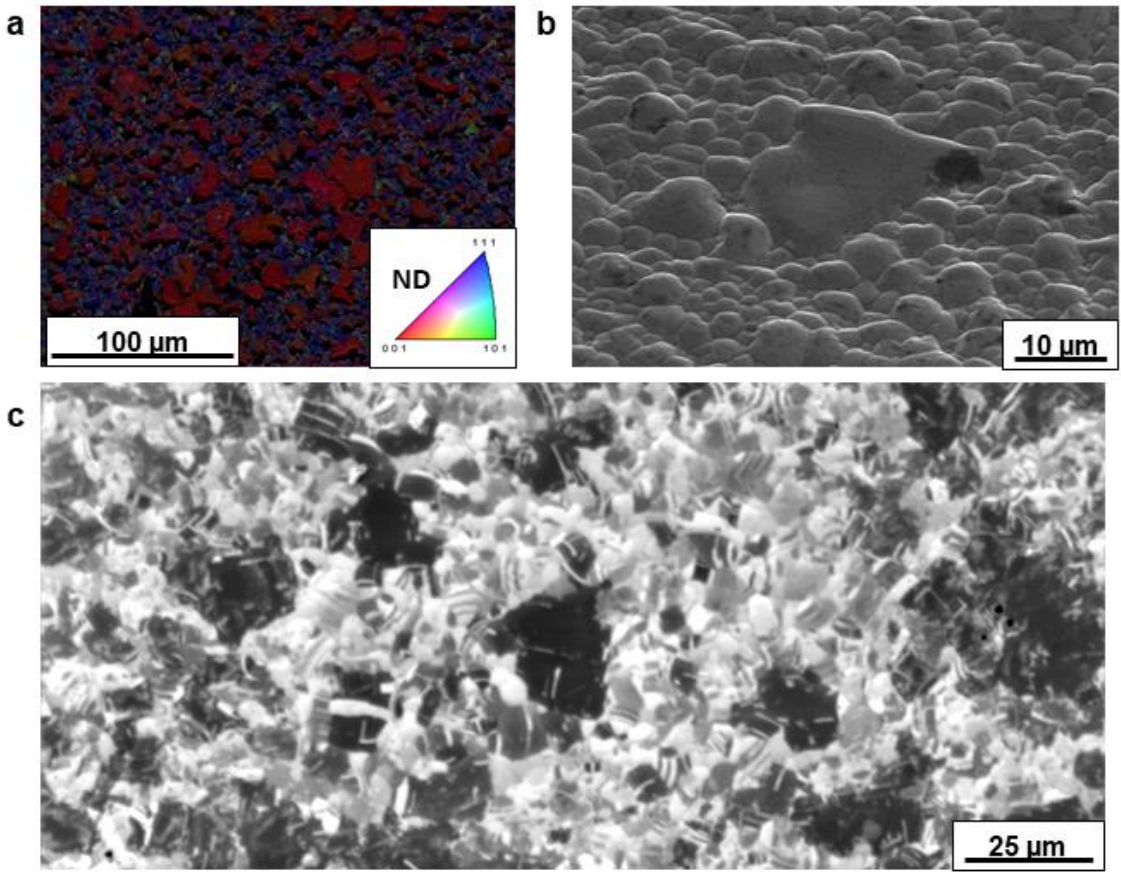

**Supplementary Figure 1 –** Ag thin film following annealing at 600°C/5 h. (a) EBSD map showing the coarser grains as <100> || ND (normal direction). (b) <100> grains are seen as protuberance at the surface, while the finer grains exhibit lower height. (c) In the Xe-ion beam (secondary electrons detector) image from the FIB-SEM, <100> grains appear as dark. The central grain was used as lift-out for <100> single crystals.

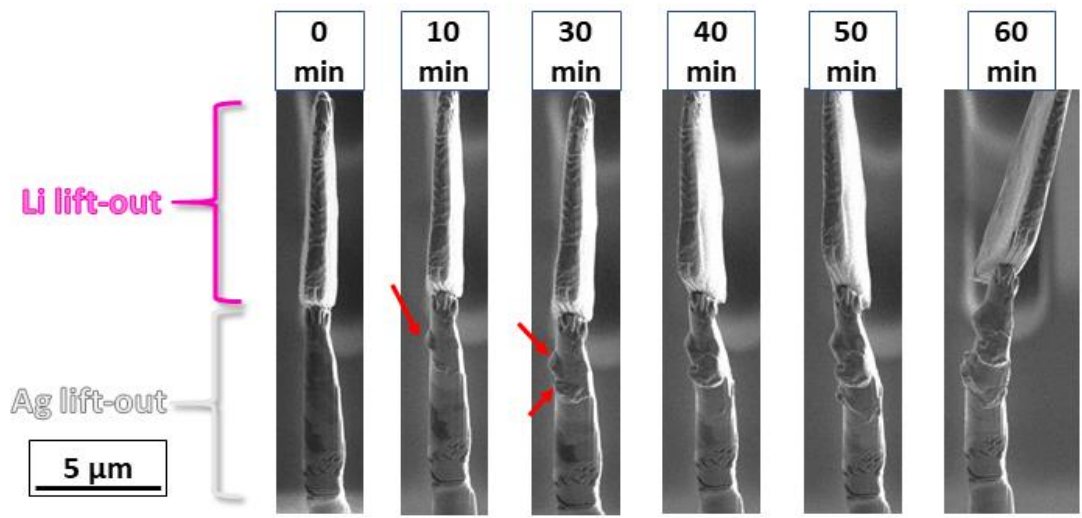

**Supplementary Figure 2 –** Li-Ag micro diffusion couple setup employing fine-grained Ag to produce atom probe tips inside the FIB. The magenta arrows highlight the swelling in the Ag substrate due to Li ingress. Some specific regions are intensively swollen, while others remain unreacted.

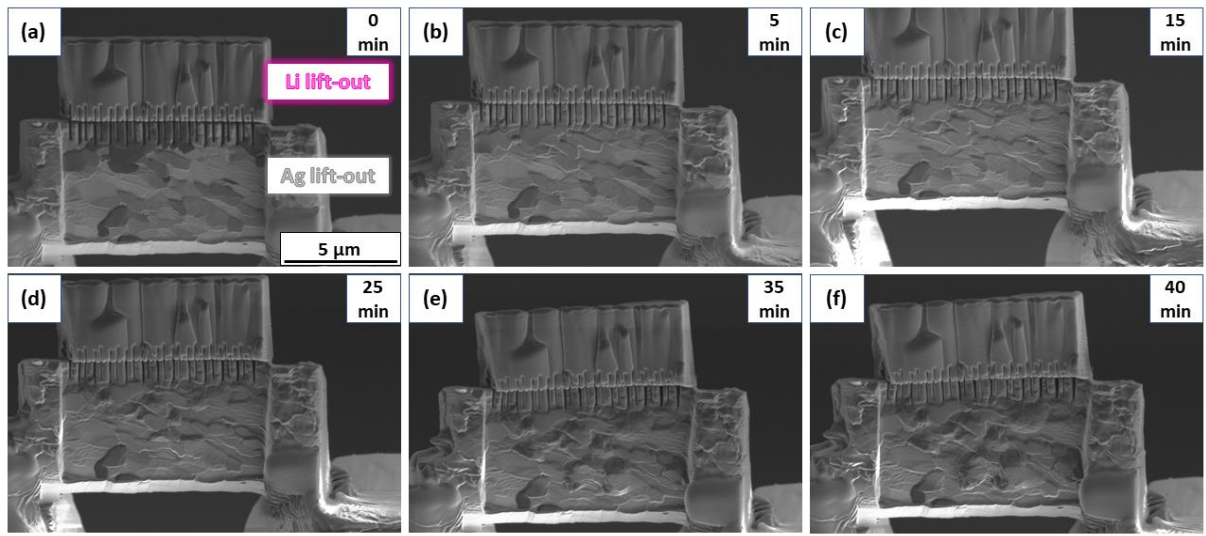

**Supplementary Figure 3 –** Ag-Li micro diffusion couple setup inside the FIB for a TEM lamella. Snapshots taken after a reaction time of (a) 0 min, (b) 5 min, (c) 15 min, (d) 25 min, (e) 35 min and (f) 40 min. Note the swelling of many specific grain boundaries and some grains.

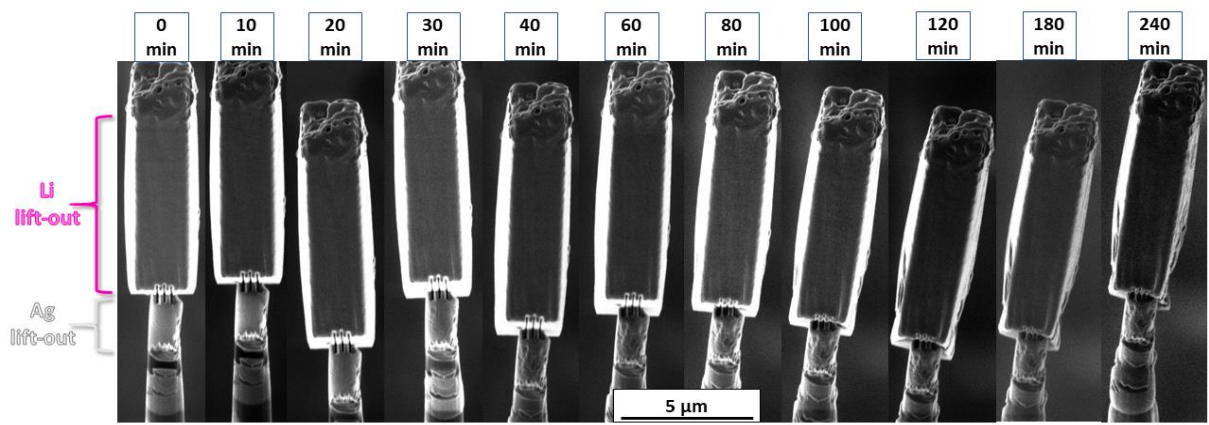

**Supplementary Figure 4 –** Ag-Li micro diffusion couple to produce APT tips from <100> Ag single crystal. The Ag single crystal does not swell, but only exhibits surface roughening, indicating Li diffusion and plating at the surface of the pillar without alloying.

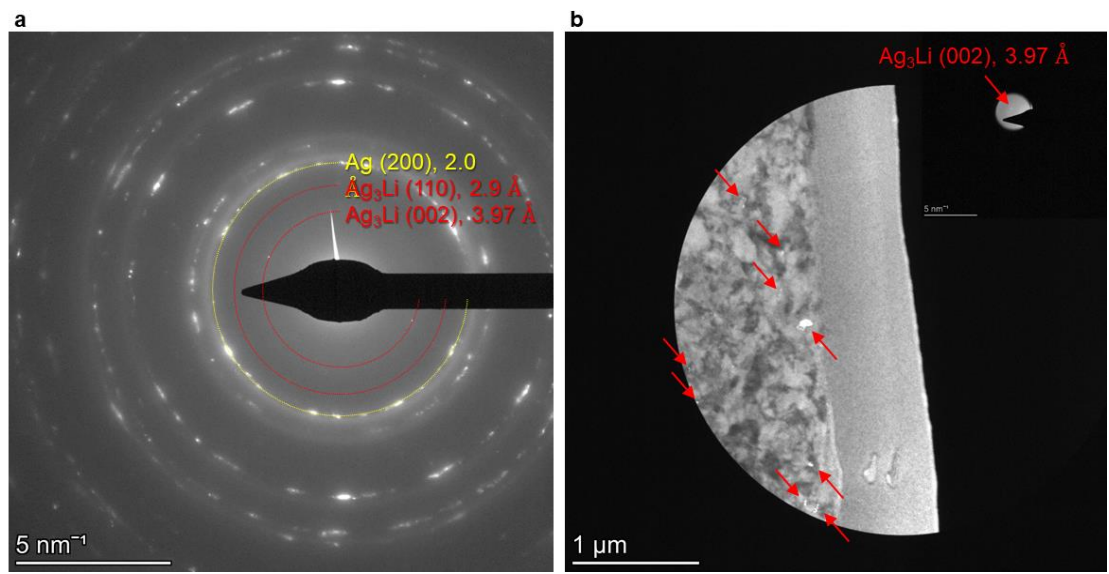

**Supplementary Figure 5 –** (a) Selected area electron diffraction pattern from the lithiated region. (b) Bright spots correspond to the isolated diffraction spot from Ag$_3$Li phase. Note that this phase appears as clusters, whereas most of the lithiated region is composed of FCC-Ag.

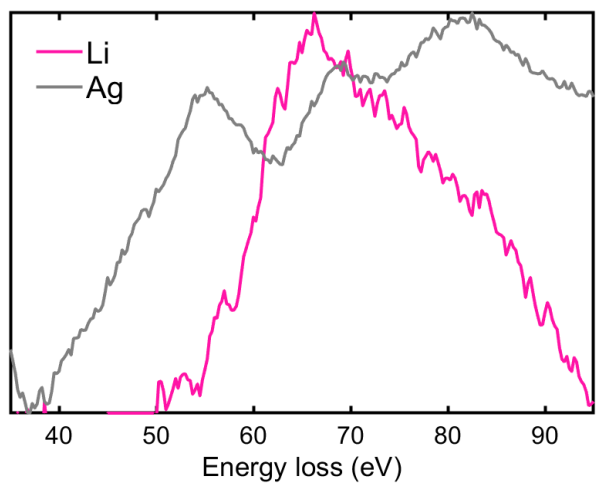

**Supplementary Figure 6 –** The electron energy loss spectrum for Li-rich and Ag-rich regions showing the Li-K and Ag-O2,3 edge from the area depicted in Figure 5d.

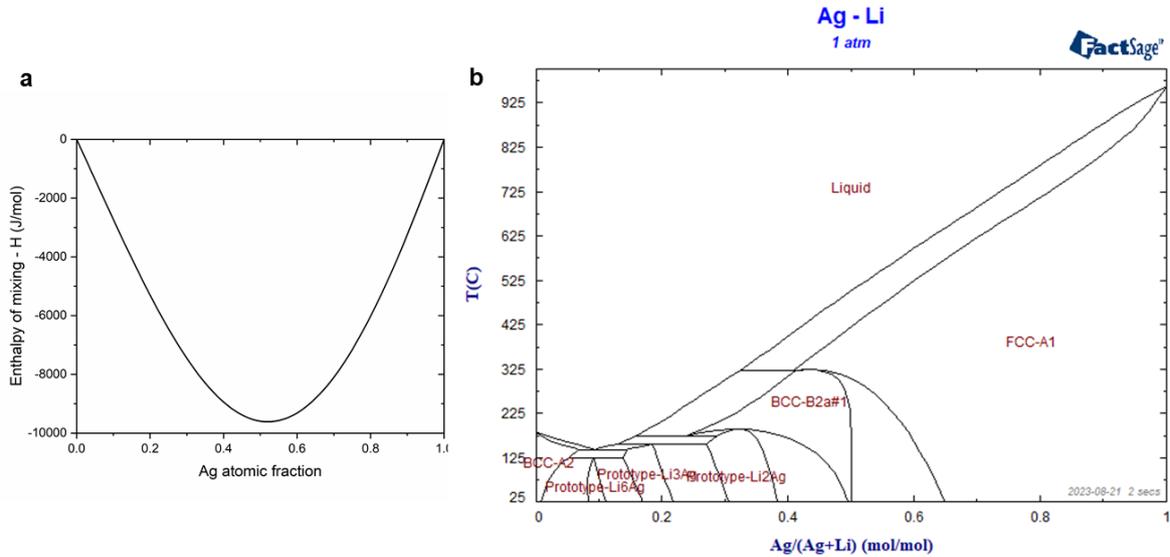

**Supplementary Figure 7 –** (a) Enthalpy of mixing for the FCC-Ag phase as function of the Ag content in the Ag-Li system. The negative enthalpy of mixing throughout the whole range indicates the tendency to form a homogeneous solid solution in the whole composition range. (b) Calculated phase diagram for the Ag-Li system in the FactSage software with the FTLite database.

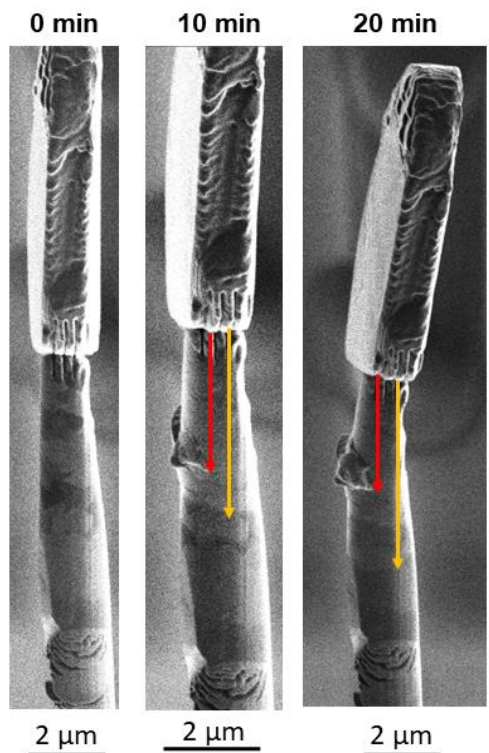

|  | Swelling (µm) | Surface contrast (µm) |
|---|---|---|
| 10 min | 3.7 | 4.77 |
| 20 min | 3.9 | 6.24 |

**Supplementary Figure 8 –** Estimation of the diffusion distance for swelling (red arrow) and surface contrast change (orange arrow).

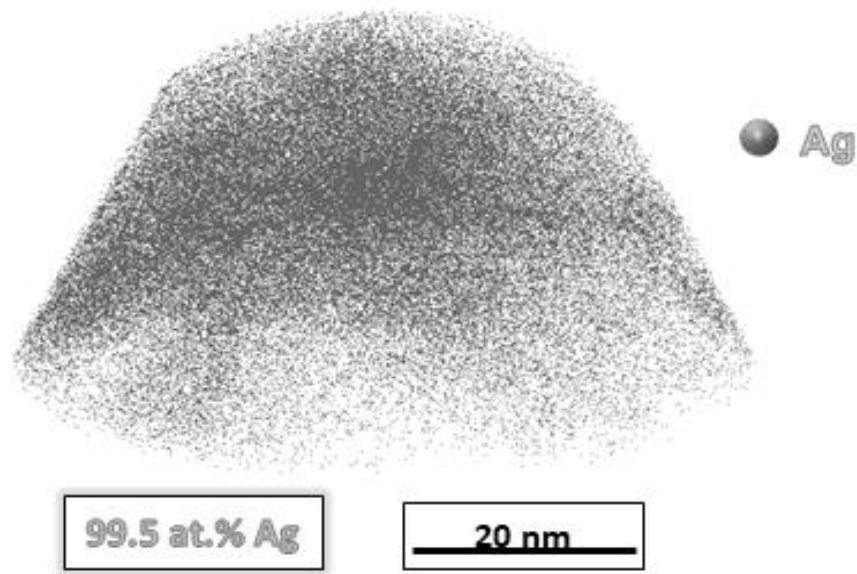

**Supplementary Figure 9 –** APT from (a) pure Ag in the <100> grain after 4 h of reaction in a micro diffusion couple tip. (b) A pure Li region at the top of another <100> grain after 4 h reaction. Only Li plating takes place at the surface of the <100> grain, with the bulk Ag remaining unreacted.

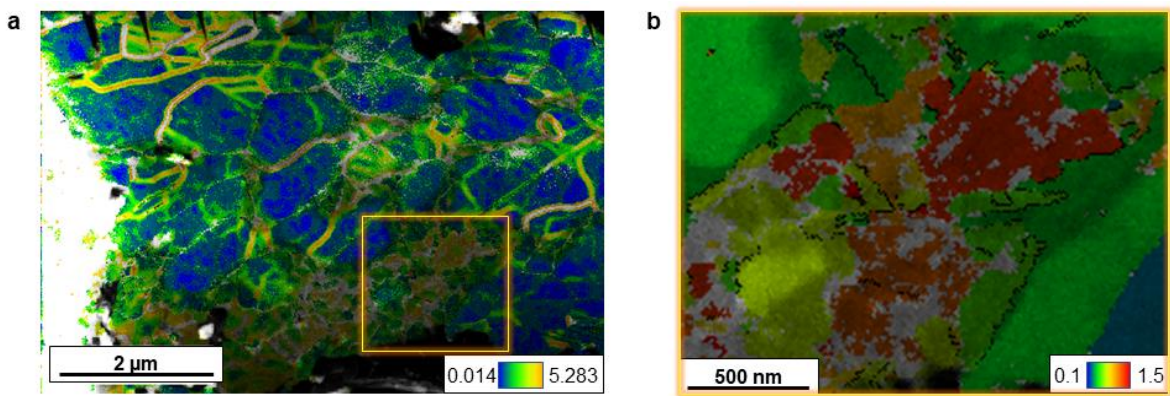

**Supplementary Figure 10 –** (a) Kernel average misorientation (KAM) and (b) grain average misorientation in color scale from the region shown by the orange box in (a), superimposed to Ag EDS signal in gray scale, obtained from the transmission Kikuchi diffraction (TKD) map. In both cases, twin boundaries are shown as black lines. In (b) note the higher KAM in the lithiated regions, indicating the presence of a higher geometrically necessary dislocation density. In (b), twinned lithiated grains exhibit lower grain average misorientation than the surrounding lithiated grains, suggesting the occurrence of dynamic recrystallization mediated by twinning.